\documentclass{evn2004}
\usepackage{graphicx}
\begin{document}
   \title{86GHz VLBI polarimetry of OVV1633+382 \\after a major mm flare}

   \author{B.W. Sohn\inst{1,2}
          \and
          T.P. Krichbaum\inst{1}
          \and
          I. Agudo\inst{1}
          \and
          A. Witzel\inst{1}
          \and
          J.A. Zensus\inst{1}
          \and
          H. Ungerechts\inst{3}
          \and
          H. Ter\"asranta\inst{4}
          }

   \institute{Max-Planck-Institut f\"ur Radioastronomie, Auf dem H\"ugel 69, 53121 Bonn, Germany
         \and
             Korea Astronomy Observatory, 61-1 Hwaam, Yuseong, Daejeon 305-348, Korea
         \and
             IRAM, Avenida Divina Pastora, 7, Nucleo Central E 18012 Granada, Spain
         \and
             Mets\"ahovi Radio Observatory, Metsahovintie, 02540 Kylm\"al\"a, Finland
             }

   \abstract{
   We report the current stage of the 86GHz VLBP monitoring program of OVV 1633+382.
    The monitoring program is still running and in this talk we will present the preliminary results
    of the first six epochs.
  }

   \maketitle
%

\section{Introduction}
The 18$^{\rm mag}$ QSO 1633+382 (4C38.41, z=1.807) showed a very pronounced outburst in 
2001/2002. With a peak amplitude of more than 9 Jy at 90GHz, this flare was
brighter than any known previous flare in this source (data available since 1980).
During onset, the mm-flare was particulary fast, with an increase of more than 2 Jy
at 230 GHz in less than 8 days. Since January 2002, the mm-flux of 1633+382 is decaying.
During this decline, however, local flux variations with amplitudes of 1-3 Jy were seen,
indicative of underlying and more rapid source activity on time scales of 1-2 months.
After the main peak occurring in 2001.99, the 90 GHz flux showed secondary maxima at
approximately half year intervals in 2002.3, 2002.7 and 2003.13. This kind of periodicity
might be explained via the lighthouse model (Camenzind \& Krockenberger 1992), which is 
based on the magnetic accelerator of Blanford \& Payne (1982). At present the millimeter
flux is nearly back to its quiescent level of 2-2.5 Jy, which the source had before the
flare began. Our VLBA Polarimetry monitoring started June 2002 during the onset of 
the flare. At cm wavelength, the flare is only marginally detected which implies
very high opacity of the source.




\section{Observations and Calibration}
\begin{table}
\caption{Observation Epochs}
\renewcommand{\arraystretch}{1.4}
\setlength\tabcolsep{5pt}
\begin{tabular}{lll}
\hline\noalign{\smallskip}
Date & Stations$^{\mathrm{a}}$ & Notes \\
\hline\noalign{\smallskip}
12 Jun 2002 & FD KP {\bf LA} NL OV MK & $^{\mathrm{b}}$ \\
28 Aug 2002 & FD {\bf KP} LA NL OV PT MK & $^{\mathrm{c}}$ \\
01 Nov 2002 & FD {\bf KP} LA NL OV PT MK & $^{\mathrm{b}}$ \\
03 Jan 2003 & FD KP {\bf LA} NL OV PT MK & \\
20 Mar 2003 & FD KP {\bf LA} NL OV PT MK & \\
23 Jun 2003 & FD KP {\bf LA} OV PT MK & \\
\hline
\end{tabular}
\label{Tab1}
\begin{list}{}{}
\item[$^{\mathrm{a}}$] bold faced characters indicate the reference antenna
\item[$^{\mathrm{b}}$] no cross pol. detection
\item[$^{\mathrm{c}}$] offset in RCP/LCP IF4 in D-terms
\end{list}
\end{table}

We observed 1633+382 at 22, 43 \& 86 GHz with the Effelsberg 100m telescope and the VLBA.
At 86 GHz, however, due to the limited frequency switching capability of Effelsberg telescope,
the source was observed only by the VLBA.  The observations were conducted with 
8 channels, 8 MHz bandwidth \& 2 bit sampling configuration. 
The data were correlated at the VLBA correlator, Socorro, NM.
For the amplitude and phase calibration, the data were read to the AIPS package.
Gain curve (GC) tables and T$_{\rm sys}$ (TY) tables were removed and reread.
VLBA procedures were used for the calibrations (e.g.
VLBAMPCL for manual phase calibration, VLBACPOL for
cross polarization phase calibration).

For hybrid imaging, the calibrated data were loaded to DIFMAP package.
At first, LL \& RR data of all sources (1633+382 and the calibrators) were 
imaged separately and compared to check if any of them is circularly polarized (i. e.
Homan \& Wardle 1999). 

After the self-calibration in DIFMAP, the data were re-read in AIPS
for the estimation of the instrumental leakage terms (Leppa\"nen et al. 1995, 
G\'omez et al. 2002). 
Obviously, the D-terms could be only well-determined when the linear polarization
(e.g. LR \& RL) of the source is prominent (a few percent of fractional polarization).
The latest three of all six epochs, show well-determined and
consistent D-terms (i.e. Fig. 1 \& Fig. 2).
Since we were not able to apply pulse-cal. tone for phase calibration at 86GHz,
there is still instrumental polarization angle offset in the data.
EVPAs of the offset uncorrected data are, however, in good agreement with
the old cm data (Cawthorne et al. 1993) and the new sub-mm data (Siringo et al. 2004).

Further EVPA test results of the calibration sources at 3mm and the
EVPA at 7mm/1.3cm will be shown.
It is still unclear why the level of cross polarization of the first three epochs (Jun. 2002
 - Nov. 2002) was much lower than the last three epochs.
We will discuss the possible instrumental and/or the intrinsic reasons.

\begin{figure}
   \centering
   \includegraphics[width=.5\textwidth]{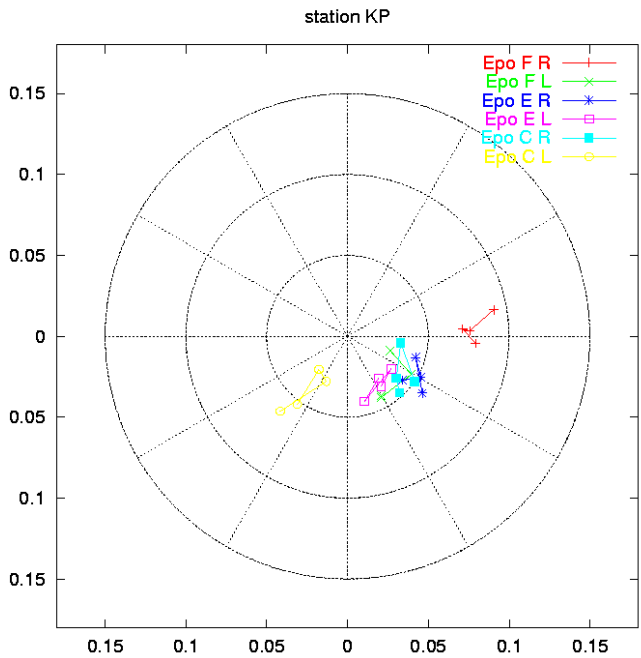}
      \caption{D-terms (amplitude and phase in polar map) of the last three epochs at 
station Kitt Peak, AZ.
Epochs were indicated as Jan. 2003 (Epo C), Mar. 2003 (Epo E), \& Jun. 2003 (Epo F). 
D-terms were determined in multi IF mode (4 IFs for each RCP \& LCP).
              }
         \label{kp}
   \end{figure}
\begin{figure}
   \centering
   \includegraphics[width=.5\textwidth]{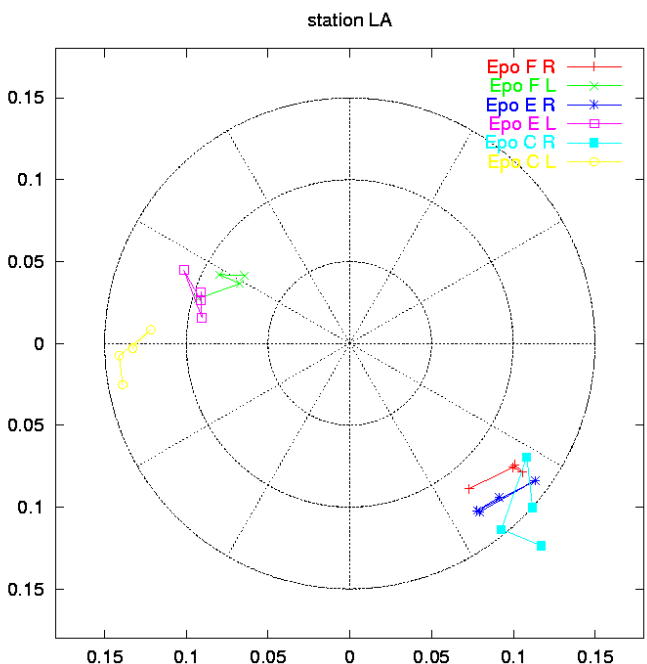}
      \caption{D-terms of the last three epochs at station Los Alamos, NM. See Fig. 1
              }
         \label{la}
   \end{figure}

\begin{figure}
   \centering
   \includegraphics[width=.5\textwidth]{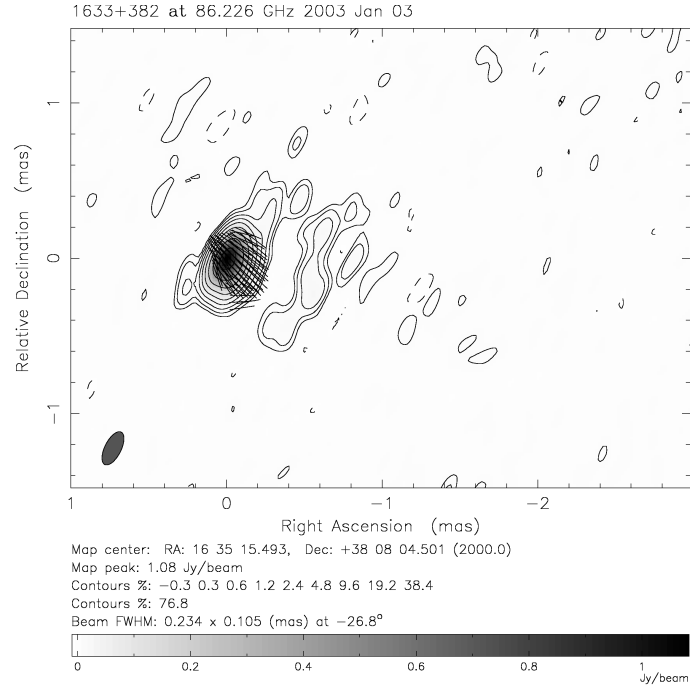}
      \caption{Total intensity map of Epoch 03 Jan. 2003 with EVPA vectors. 
The vectors are indicating EVPA. Instrumental offset of EVPA is un-corrected. 
The vector length is scaled to 5 mas/mJy with 5mJy/beam cutoff.
              }
         \label{EPO_C}
   \end{figure}

\begin{figure}
   \centering
   \includegraphics[width=.5\textwidth]{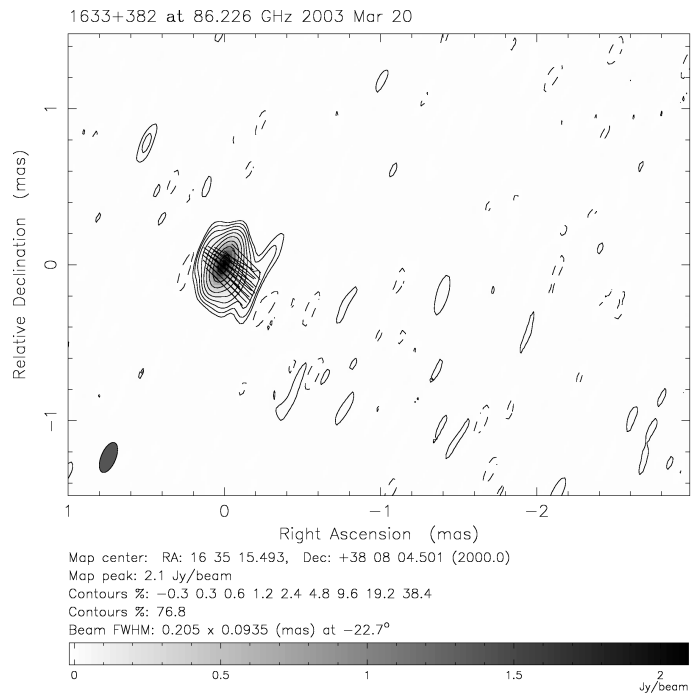}
      \caption{Total intensity map of Epoch 20 Mar. 2003 with EVPA vectors. See Fig.3
              }
         \label{EPO_E}
   \end{figure}
\begin{figure}
   \centering
   \includegraphics[width=.5\textwidth]{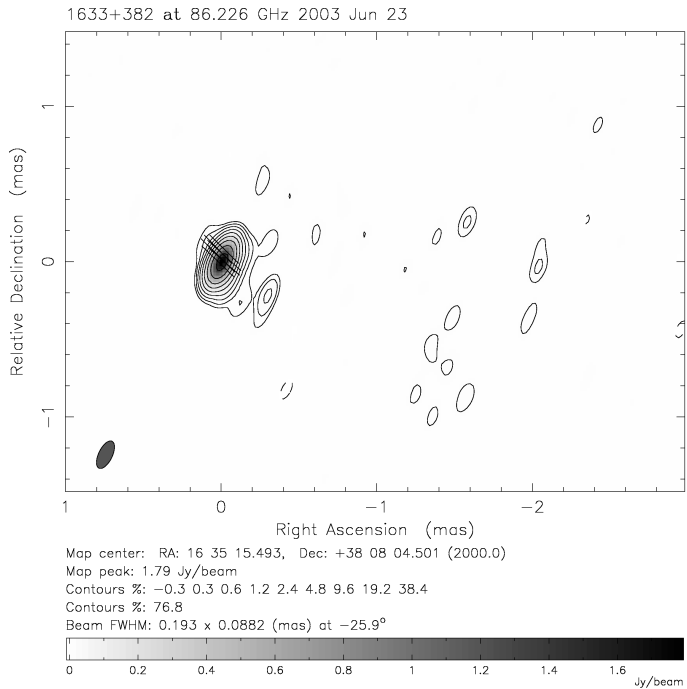}
      \caption{Total intensity map of Epoch 23 Jun. 2003 with EVPA vectors. See Fig.3
              }
         \label{EPO_F}
   \end{figure}

%

\cleardoublepage

\begin{thebibliography}{}
   \bibitem[1993]{cawthorne} Cawthorne, T. V., Wardle F. C., Roberts, D. H., et al. 1993,
      ApJ, 416, 496

   \bibitem[2002]{gomez} G\'omez, J. L., Marscher, A. P., Alberdi, A., et al. 2002,
      VLBA Scientific Memo No. 30, NRAO
   
   \bibitem[1999]{homan} Homan, D. C., \& Wardle F. C. 1999,
      ApJ, 556, 113

   \bibitem[1995]{leppa} Lepp\"anen, K. J., Zensus, J. A., \& Diamond, P. J. 1995,
      AJ, 110, 2479
   

   \bibitem[2004]{siringo} Siringo, E. et al. 2004,
      A\&A, 422, 751

\end{thebibliography}
\end{document}